\documentclass[aps,onecolumn]{revtex4}
\usepackage{amsfonts}
\usepackage{amsmath}
\usepackage{amssymb,epsf}
\usepackage{color}
\usepackage{hyperref}

\begin{document}

\title{Three dimensional Lifshitz-like black hole in a special class of $F(R)$ gravity}
\author{S. H. Hendi$^{1,2}$\footnote{
email address: hendi@shirazu.ac.ir}, R. Ramezani-Arani$^{3}$, E.
Rahimi$^{3}$} \affiliation{$^1$Physics Department and Biruni
Observatory, College of Sciences, Shiraz
University, Shiraz 71454, Iran\\
$^2$Canadian Quantum Research Center 204-3002 32 Ave Vernon, BC V1T 2L7 Canada\\
$^{3}$Department of Elementary Particles, Faculty of Physics,
University of Kashan, Kashan, Iran}

\begin{abstract}
Regarding a special class of pure $F(R)$ gravity in three dimensions, we
obtain, analytically, Lifshitz-like black hole solutions. We check the
geometrical properties of the solutions which behave such as charged BTZ
black holes in special limit. We also investigate the thermodynamic
properties of the solutions and examine the first law of thermodynamics and
Smarr formula. In addition, we study thermal stability via the heat capacity
and discuss the possibility of criticality in the extended phase space.
\end{abstract}

\maketitle

\section{Introduction}

The generalization of Einstein's Lagrangian to a more general invariant of
the Riemann tensor, an arbitrary function of the Ricci scalar, is considered
by Buchdahl in $1970$ \cite{Buchdahl}. Unlike Einstein's gravity, $F(R)$
theory can explain the accelerated expansion \cite%
{Perlmutter,Riess,Riess1} and structure formation of the Universe
without considering dark energy or dark matter. Such a theory may
avoid the known instability \cite{Woodard} and is coincident with
Newtonian and post-Newtonian approximations \cite{Capozziello,
Stabile}. So, different solutions of $F(R)=R+f(R)$ gravity with
various motivations have been
proposed and their properties are investigated \cite%
{Odintsov1,Starobinsky, Bamba,Duvvuri,Panah,Cognola}. Regardless
of the trivial spherical symmetric solutions of $F(R)$ models of
gravity, analyzing new solutions of this theory with nontrivial
topologies are interesting.

Einstein's gravity cannot explain the non-relativistic scale-invariant
theory, and it should be regarded as an effective theory that does not show
Galilian symmetry and brakes down at some scales. In order to overcome such
a problem, one can use of Horava-Lifshitz \cite{Horava,Horava1} approach
introducing a metric which shows an anisotropic scale invariant between time
and space
\begin{equation}
t\longrightarrow \lambda ^{z}t\text{ \ \ \ \ \ \ \ \ \ \ \ \ \ \ \ \ \ \ \ }%
x\rightarrow \lambda x,  \label{crit}
\end{equation}
where $z$ is called as dynamic critical exponent of the Horava-Lifshitz
theory. This theory provides a violation of the Lorentz symmetry in high
energies which is a possible characteristic of quantum gravity theory.

The static vacuum solutions of a Lifshitz model in ($2+1$)-dimensions has
been investigated in \cite{Shu}. Three dimensional black hole solution
introduced by Banados-Teitelboim-Zanelli (BTZ) \cite{Banados} is one of the
interesting subjects for gravitating systems in recent years \cite%
{Anabal,Hodgkinson,Moon,Hassaine,Xu,Wu}. These black holes have
been used to develop the knowledge of gravitational interaction in
low dimensional manifolds and also improvement in quantum theory
of gravity, string theory and gauge field theory
\cite{Witten,Carlip}. The holography of the BTZ black holes has
been investigated in details \cite{Fuente}. In addition,
thermodynamic properties of BTZ black holes have been studied
during the last years \cite{Panahiyan,Eslam Panah,Liang}.

Thermodynamical structure of the black holes has been of great interest \cite%
{Wald,Wald1}. Especially in recent years, considering the
cosmological constant as a thermodynamical variable and working in
the extended phase space lead to find additional analogy between
the black holes and the behavior of the van der Waals liquid/gas
system \cite{Gibbons,Bret,Kastor,Huan,Hendiii}.

The phase transition of a black hole plays an important role in
exploring its critical behavior near the critical point. In order
to study the phase transition, one may adopt different approaches
\cite{Ma,Mo,Zhang}. One of these methods is studying the behavior
of the heat capacity. It is argued that divergence points of the
heat capacity hint us the existence of phase transition. In
addition, it is notable that a thermally stable black hole has a
non-negative heat capacity. In addition, focusing on the Gibbs
free energy and its derivatives, one can find the possibility of
the phase transition and its order. Moreover, working on the
extended phase space, one can extract some useful information
through the $PV$ and $PT$ diagrams. The stability of BTZ black
hole has been studied in some papers \cite{Myung}. Here, we are
going to evaluate thermal stability of the Lifshitz black holes in
three dimensional $F(R)$ gravity with constant Ricci scalar.

The structure of this paper is as follows: At first, we obtain three
dimensional Lifshitz-like black hole in the context of a special class of
the $F(R)$ gravity and investigate its geometric and thermodynamic
properties. Then, we work in the extended phase space thermodynamics and
examine thermal stability. We also show that for the obtained Lifshitz-like
solutions there is no critical behavior. In addition, we check the validity
of the first law of thermodynamics and the Smarr relation, and find that the
Lifshitz parameter modifies the conserved charges. Final section is devoted
to concluding remarks.

\section{exact solutions of $F(R)$ gravity in three dimensions}

In order to study $3-$dimensional black holes in pure $F(R)$ gravity, we
employ the following action
\begin{equation}
S=\int_{\mathcal{M}}d^{2+1}x\sqrt{-g}F(R),  \label{Action}
\end{equation}%
where $\mathcal{M}$ is a three-dimensional bulk manifold. In this action, $R$
and $F(R)$ are, respectively, the Ricci scalar which we regard it as a
constant ($R_{0}$) and an arbitrary function of it. It is note that the
action is constructed as a pure geometric (gravitational) theory without
matter field.

Using the variational principle, it is straightforward to obtain the
following field equation
\begin{equation}
G_{\mu \nu}F_{R}-\frac{1}{2}g_{\mu \nu}[F(R)-RF_{R}]-[\nabla _{\mu }\nabla
_{\nu }-g_{\mu \nu}\square]F_{R}=0,  \label{FE2}
\end{equation}
where $F_{R}=\frac{dF(R)}{dR}$.

Hereafter, we follow the method of Ref. \cite{Calza}, which is
applicable for the special class of $F(R)$-gravity models
satisfying two constraints, simultaneously, $F(R_{0})=0$ and
$F_{R}=0$. Regarding the mentioned constraints, we find that the
vacuum field equation (\ref{FE2}) are automatically satisfied with
arbitrary Ricci scalar $R_{0}$. It is notable that the mentioned
class of $F(R)$-theory does not lead to the usual general
relativity since the vacuum field equation of general relativity
identically
satisfied $F_{R}=1$ with vanishing Ricci scalar. As it is mentioned in \cite%
{Calza}, there are several models for the early-time inflation or late-time
accelerated expansion that can satisfy the mentioned constraints.

Here, our main motivation is to study the thermodynamic aspects of black
hole solutions in a Lifshitz-like background spacetime. Therefore, we
consider the metric of $3$-dimensional spacetime as

\begin{equation}
ds^{2}=-(\frac{r}{r_{0}})^{z}B(r)dt^{2}+\frac{dr^{2}}{B(r)}+r^{2}d\phi ^{2}.
\label{Metric2}
\end{equation}%
where $z$ is a real number so called Lifshitz -like parameter and $r_{0}$ is
an arbitrary (positive) length scale. Considering the mentioned metric, we
can extract the metric function for $R=R_{0}$, where
\begin{equation}
R_{0}=-B^{\prime \prime }-\frac{3z+4}{2r}B^{\prime }-\frac{z^{2}}{2r^{2}}B,
\label{RicciS}
\end{equation}%
where we used the usual notation $B=B(r)$, $B^{\prime }=\frac{dB(r)}{dr}$
and $B^{\prime \prime }=\frac{d^{2}B(r)}{dr^{2}}$ for the sake of brevity.
Considering Eq. (\ref{RicciS}), we can find the following exact solution
\begin{equation}
B(r)=-\frac{m}{r^{\gamma }}+\frac{q^{2}}{r^{\delta }}-\Lambda r^{2},
\label{B(r)}
\end{equation}%
in which $m$ and $q$ are two integration constants of the second order
differential equation and $\Lambda $ is (positive/negative or zero) constant
that its value is depending on the sign/value of $R_{0}$ as
\begin{equation*}
\Lambda =\frac{2R_{0}}{z^{2}+6z+12}.
\end{equation*}

In addition, $m$ and $q$ are two integration constants while $\gamma $ and $%
\delta $ are defined as
\begin{eqnarray}
\gamma &=&\frac{1}{4}\left( 3z+2-\sqrt{z^{2}+12z+4}\right)  \label{b} \\
\delta &=&\frac{1}{4}\left( 3z+2+\sqrt{z^{2}+12z+4}\right) .
\end{eqnarray}

%%%%%%%%%%%%%%%%%%%%%%%%%%%%%%%%%%%%%%%%%%%%%%%%%%%
\begin{figure}[tbp]
$%
\begin{array}{cc}
\epsfxsize=8cm \epsffile{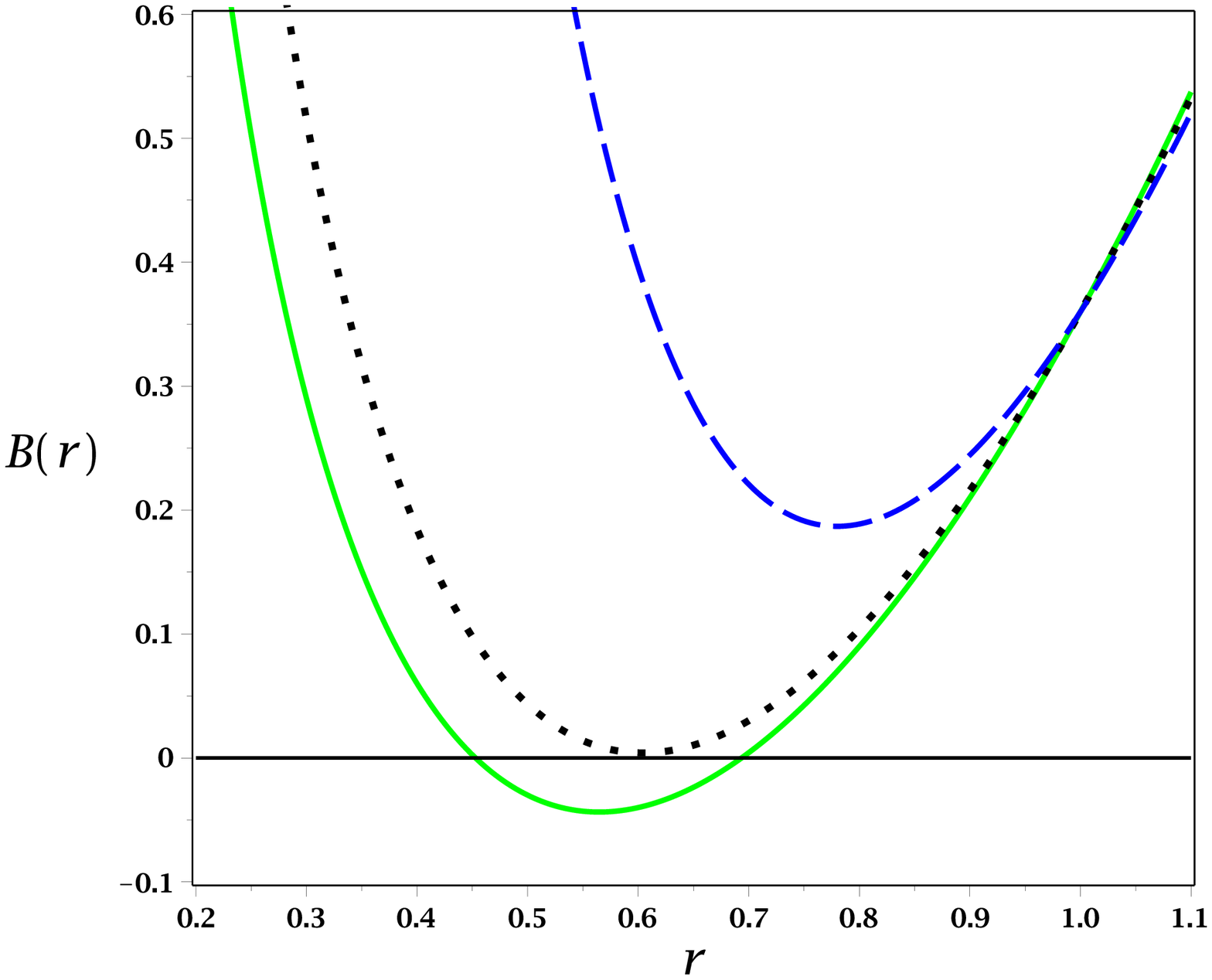} & \epsfxsize=8cm
\epsffile{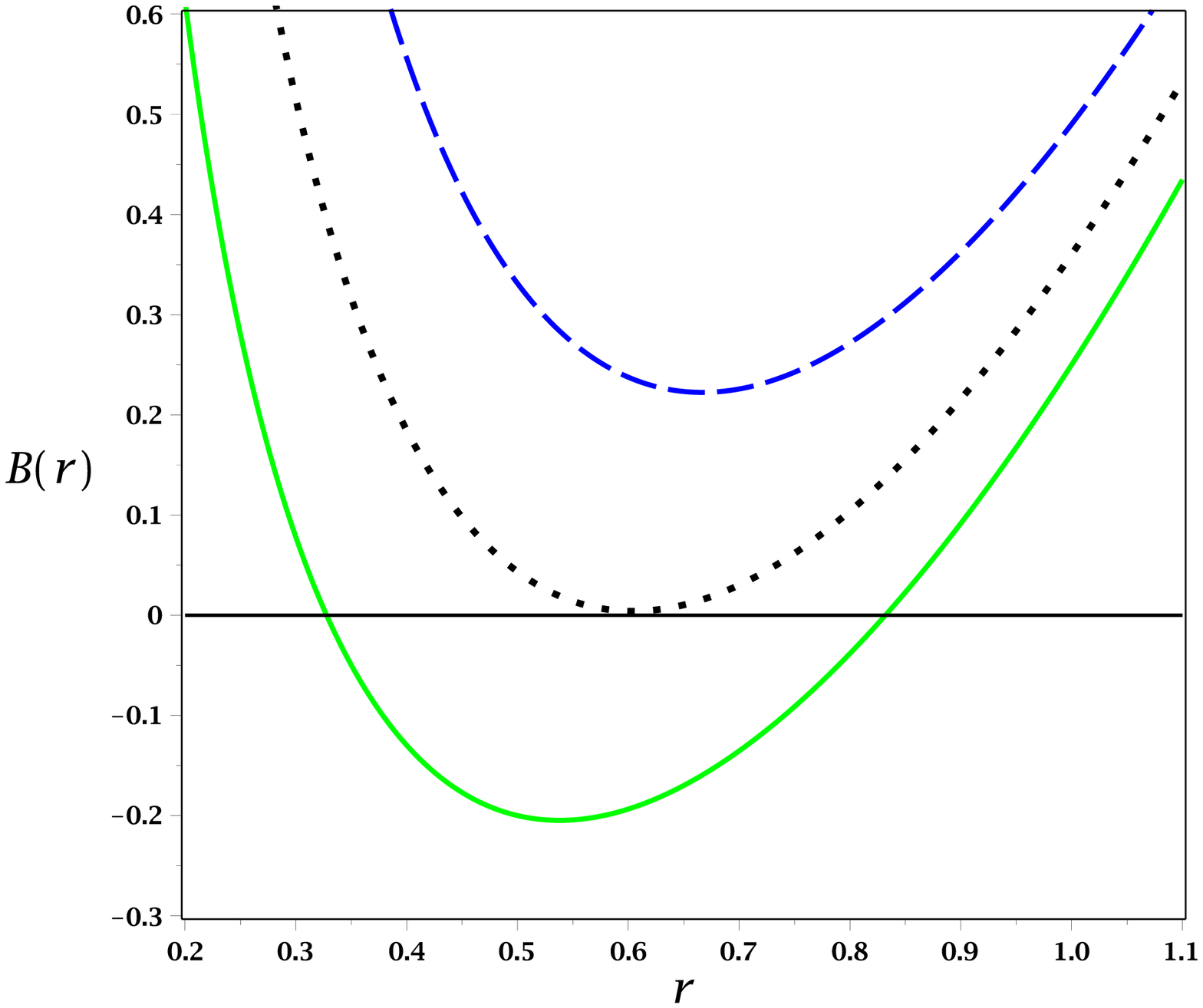}  \\
\epsfxsize=8cm \epsffile{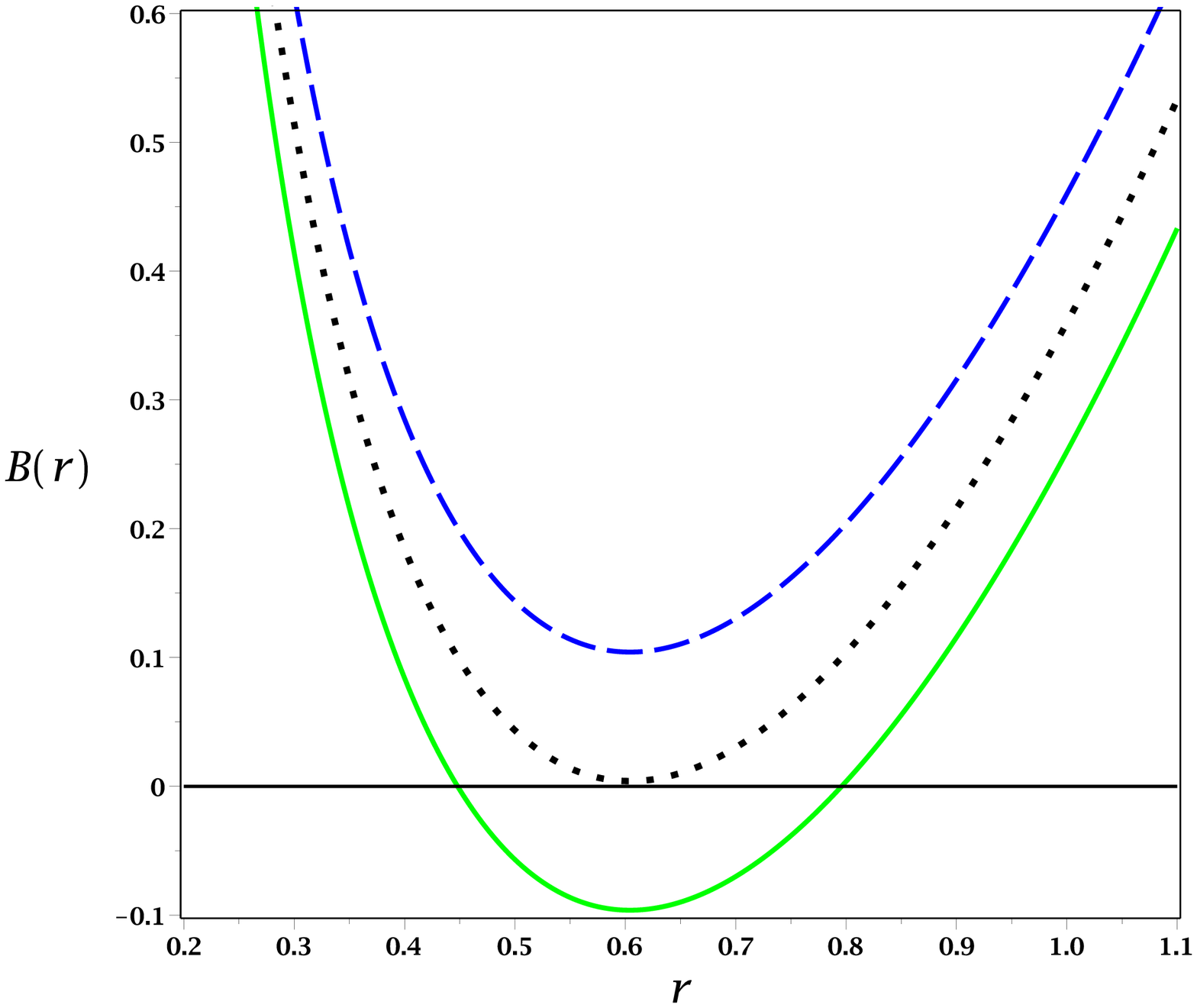} & \epsfxsize=8cm
\epsffile{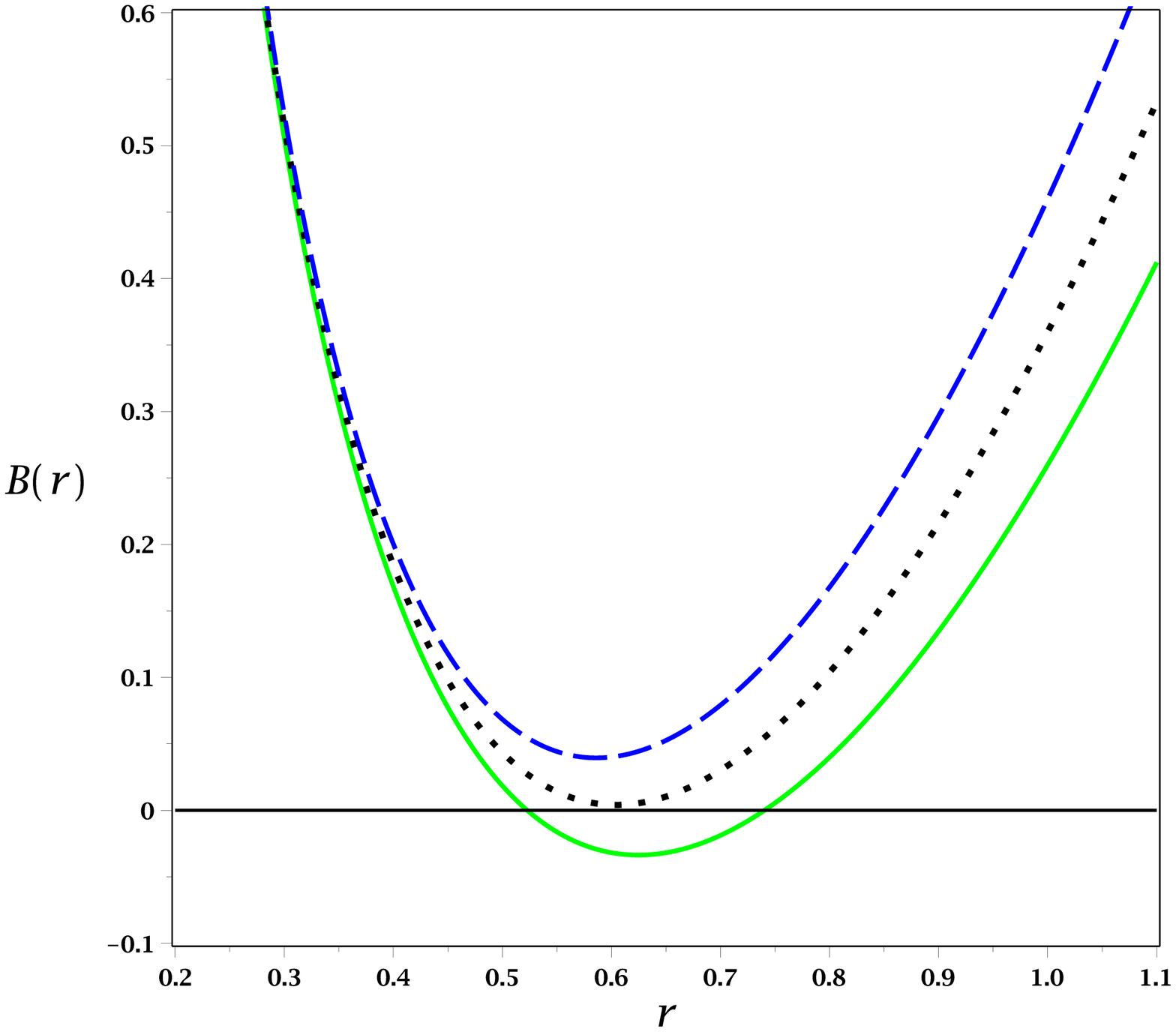}
\end{array}
$%
\caption{ $B(r)$ versus $r$. \\ \textbf{Left-upper panel:}
$q=0.6$, $\Lambda=-1$, $m=1$, and $z=0$ (continuous line), $z=0.1$
(dotted line) and $z=1$ (dashed line). \\ \textbf{Right-upper
panel:} $z=0.1$, $\Lambda=-1$, $m=1$, and $q=0.5$ (continuous
line),
$q=0.6$ (dotted line) and $q=0.7$ (dashed line). \\
\textbf{Left-lower panel:} $q=0.6$, $\Lambda=-1$, $z=0.1$, and
$m=1.1$ (continuous line), $m=1$ (dotted line) and $m=0.9$ (dashed
line).
\\ \textbf{Right-lower panel:} $z=0.1$, $q=0.6$, $m=1$, and
$\Lambda=-0.9$ (continuous line), $\Lambda=-1$ (dotted line) and
$\Lambda=-1.1$ (dashed line).} \label{Figmetric}
\end{figure}
%%%%%%%%%%%%%%%%%%%%%%%%%%%%%%%%%%%%%%%%%%%%%%%%%%%

In order to interpret the solutions as black holes, we should examine the
existence of horizon and singularity for the singular black holes. The
presence of singularity could be investigated by studying curvature scalars
for which we choose the Kretschmann scalar. It is a matter of calculation to
show that for these solutions, the Kretschmann scalar is
\begin{equation}
R_{\alpha \beta \gamma \delta }R^{\alpha \beta \gamma \delta }
=B^{\prime \prime 2}+\frac{z}{r^{2}}[3rB^{\prime
}+(z-2)B]B^{\prime \prime } +\frac{9z^{2}+2}{4r^{2}}B^{\prime 2}+\frac{z(3z^{2}-6z+4)}{2r^{3}}%
BB^{\prime }  +\frac{z^{2}\left( z^{2}-4z+8\right) }{4r^{4}}B^{2}.
\label{RR}
\end{equation}

Regarding Eq. (\ref{RR}) with the obtained metric function, we
find that Kretschmann scalar diverges at $r=0$ and is finite for
$r\neq 0$. In addition, according to the Fig. \ref{Figmetric}, one
finds the metric function has at least one real positive root
(with positive slope), and therefore, the mentioned solutions can
be interpreted as black holes. In other words, one finds that
depending on the values of parameters ($z$, $m$, $\Lambda$ and
$q$), the metric function has two real positive roots ($r_{-}$ and
$r_{+}$), one degenerate root
($r_{ext}=-\frac{(\gamma+2)\Lambda}{(\delta-\gamma)q^2} $) or it
may be positive definite. Therefore, these solutions may be
interpreted as black holes with two horizons (one Cauchy horizon
at $r_{-}$ and one event horizon at $r_{+}$), extreme black holes
($B(r)|_{r=r_{ext}}=B'(r)|_{r=r_{ext}}=0$) or naked singularity
(see all panels of Fig. \ref{Figmetric} for more details).

Regarding $z=0$, one may expect to obtain the usual BTZ black hole
solutions. However in this limit the metric function reduces to $%
B(r)=-\Lambda r^{2}-M+\frac{q^{2}}{r}$. The first two terms are the usual
cosmological constant and mass term while third term differs from the
charge-term of BTZ-Maxwell solutions which is logarithmic form. Nonetheless,
we find that the last term can be interpreted as a charge-term in
power-Maxwell nonlinear electrodynamics with the following total charge%
\begin{equation}
Q=\frac{3}{16}\frac{2^{\frac{3}{4}}}{r_{0}^{\delta -1}}\sqrt{q}.  \label{Q}
\end{equation}%
In other words, such a metric function is completely in agreement with that
of charged black hole solution of Einstein-power Maxwell invariant
(Einstein-PMI) gravity when the nonlinearity parameter is chosen $%
s=dimension/4$ which is conformally invariant Maxwell source. This
result implies an interesting result. Indeed, $F(R)$ gravity
provides a framework for putting the gravity and PMI nonlinear
electrodynamics in a unified context through the pure geometry. It
is notable that the same approach is applied in four dimensional
$F(R)$ gravity \cite{H-R-R-4} and its solution is in agreement
with the above statement ($s=1$ since $dimension=4$). The general
form of $d-$dimensional solutions is addressed in the appendix.

\section{Thermodynamic behavior and thermal stability}

Regarding an exact solution of gravitational field equation, we have to
check the stability of the solutions. There are two main stability criteria
which a known as dynamic and thermodynamic stability. In this paper we
investigate the latter one. Considering the black hole as a thermodynamic
system, one has to examine the validity of the first law and thermal
stability. To do so, we should calculate thermodynamic and conserved
quantities.

\subsection{Conserved and thermodynamic quantities}

Since the employed metric contains a temporal Killing vector ($\partial
/\partial t$), we can use the concept of surface gravity ($\kappa $) to
calculate the temperature of black holes at the event horizon $r_{+}$
\begin{equation}
T=\frac{\kappa }{2\pi }=\left. {\ \frac{B^{\prime }(r)}{4\pi }\left( \frac{r%
}{r_{0}}\right) ^{\frac{z}{2}}}\right\vert _{r=r_{+}}.  \label{(Temp)}
\end{equation}%
Regarding Eq. (\ref{(Temp)}) with the metric function (\ref{B(r)}), one can
find
\begin{equation}
T=\frac{\left( \frac{r_{+}}{r_{0}}\right) ^{\frac{z}{2}}}{4\pi }\left[ \frac{%
q^{2}(\gamma -\delta )}{r_{+}^{1+\delta }}-(2+\gamma )\Lambda r_{+}\right] ,
\label{Temp2}
\end{equation}%
where $r_{+}$ is the radius of event horizon determined from $B(r_{+})=0$
due to the fact that the metric function vanishes on the event horizon.

Comparing the solutions with BTZ black holes or using the
Ashtekar-Magnon-Das (AMD) formula for a far away observer, we find
that the finite total mass can be written as
\begin{equation}
M=\frac{m_{0}}{8}=\frac{m}{8}r_{0}^{-\gamma }.  \label{Mass1}
\end{equation}

Here, we desire to examine the validity of the first law of thermodynamics.
Evaluating the metric function on the event horizon $(B(r_{+})=0)$, one can
obtain the geometrical mass, $m_{0}$, as a function of $r_{+}$. Inserting $%
m_{0}(r_{+})$ in Eq. (\ref{Mass1}), one finds
\begin{equation}
M=\frac{1}{8}\left( -\Lambda r_{+}^{2}-\frac{q^{2}}{r_{+}^{\delta }}\right)
\left( \frac{r_{+}}{r_{0}}\right) ^{\gamma }.  \label{Mass}
\end{equation}

Since we consider a special class of $F(R)$ gravity with $F_{R}=0$,
calculation of the entropy based on Wald's formula is problematic. In order
to obtain the entropy of black holes in such a class of $F(R)$ gravity, one
can respect the validity of the first law of thermodynamics. So we can
obtain the entropy as
\begin{equation*}
\delta S=\frac{1}{T}\delta M,
\end{equation*}%
and therefore, it is a matter of calculation to show that the following
relation holds
\begin{equation}
S=\int \frac{dM}{T}=\frac{\pi r_{+}}{2\gamma -z+2}\left( \frac{r_{0}}{r_{+}}%
\right) ^{\frac{z}{2}-\gamma }.  \label{(Entropy)}
\end{equation}

It is notable that the obtained relation for the entropy reduces to the area
law for $z=0$ ($\gamma =0)$.

\subsection{Examine thermal stability and phase transition}

Regarding the variation of the cosmological constant as the vacuum
expectation value of a quantum field, one may expect to consider
it and its conjugate in the first law of thermodynamics. It is
notable that the finite mass of black holes interpreted as the
enthalpy ($M \equiv H$) of the system rather than the internal
energy in the extended phase space \cite{Dolan}
\begin{eqnarray}
P &=&-\frac{\Lambda }{8\pi },  \label{lambda} \\
V &=& \left( \frac{\partial M}{\partial P} \right)_{S,q}=\pi r_{+}^{2}\left(
\frac{r_{+}}{r_{0}}\right) ^{\gamma }.
\end{eqnarray}

Regarding the relation of temperature (\ref{Temp2}) with the mentioned
equation of pressure (\ref{lambda}), we can obtain the so-called equation of
state
\begin{equation}
P=\frac{1}{2r_{+}^{2}(\gamma +2)}\left[ \frac{Tr_{+}}{\left( \frac{r_{+}}{%
r_{0}}\right) ^{\frac{z}{2}}}+\frac{q^{2}(\delta -\gamma )}{4\pi
r_{+}^{\delta }}\right] .  \label{pressur}
\end{equation}

\bigskip To obtain the critical point, we use the feature of the inflection
point of the $P-V$ diagram at the critical point. In other words, the first
and second derivatives of the pressure with respect to the volume vanish at
the critical point, i.e.:\bigskip
\begin{eqnarray}
\left( \frac{\partial P}{\partial r_{+}}\right) _{T} &=&0,  \label{dP} \\
\left( \frac{\partial ^{2}P}{\partial r_{+}^{2}}\right) _{T} &=&0,
\label{ddP}
\end{eqnarray}%
in which after some calculations, we find
\begin{equation}
\frac{q^{2}(\gamma -\delta )(\delta +2)}{(\frac{r_{0}}{r_{+}})^{\frac{z}{2}%
}r_{+}^{1+\delta }}-2\pi T(z+2)=0,  \label{drive}
\end{equation}%
\begin{equation}
\frac{q^{2}(\delta -\gamma )(\delta +2)(\delta +3)}{(\frac{r_{0}}{r_{+}})^{%
\frac{z}{2}}r_{+}^{1+\delta }}+\pi T(z+2)(z+4)=0.  \label{drive1}
\end{equation}

According to the equations (\ref{drive}) and (\ref{drive1}) , it is clear
that these equations cannot admit any critical point for a positive real
value of $r_{+}$. To conclude, for the Lifshitz-like black holes in three
dimensions addressed in this paper, we could not observe any van der Waals
like behavior. So it seems that obtained solutions may be thermally stable.
To confirm this statement, we can calculate the heat capacity as%
\begin{equation}
C_{P,Q}=T\left( \frac{\partial S}{\partial T}\right) _{P,Q}=\frac{\pi r_{+}}{%
(\frac{r_{+}}{r_{0}})^{\frac{z}{2}-\gamma }}\frac{\mathcal{A}}{\mathcal{B}},
\label{CQ}
\end{equation}%
where%
\begin{eqnarray}
\mathcal{A} &=&r_{+}^{\delta +2}P-\frac{(\delta -\gamma )q^{2}}{8\pi (\gamma
+2)}, \\
\mathcal{B} &=&(z+2)r_{+}^{2+\delta }P+\frac{(\delta -\gamma )(2\delta
+2-z)q^{2}}{8\pi (\gamma +2)}.
\end{eqnarray}

\begin{figure}[tbp]
$%
\begin{array}{c}
\epsfxsize=8cm \epsffile{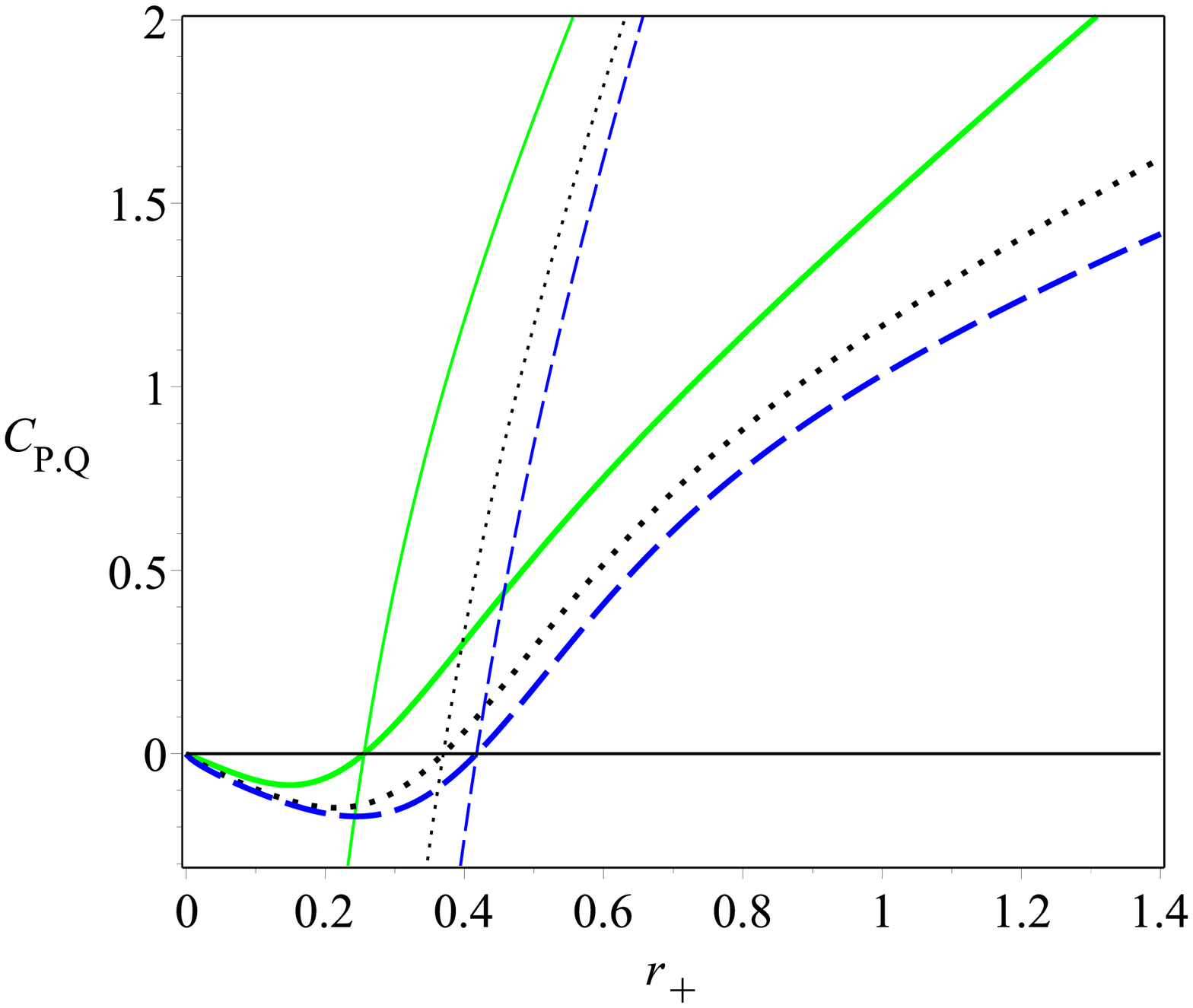} \\
\epsfxsize=8cm \epsffile{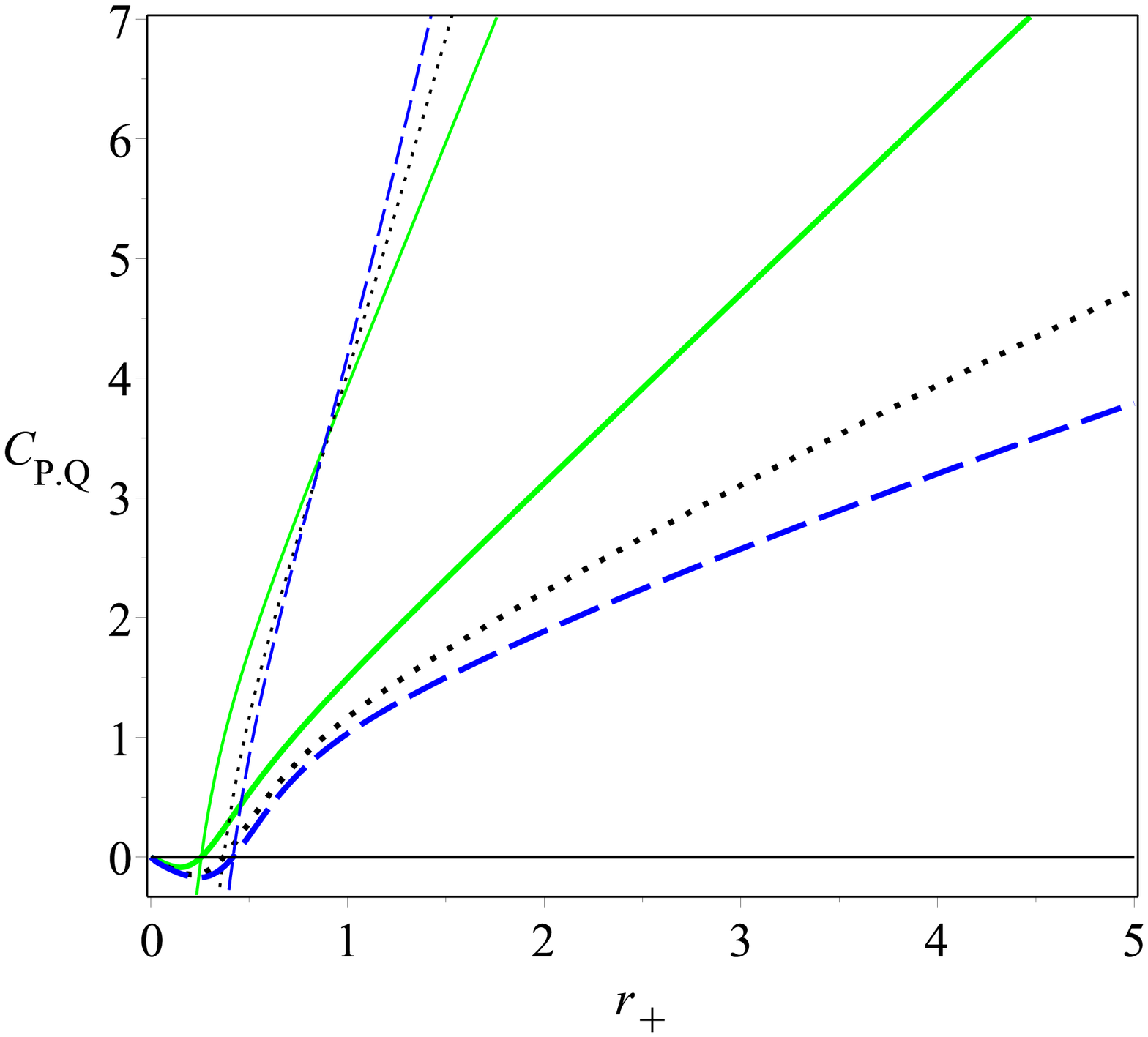}%
\end{array}
$%
\caption{ Different scales: Heat Capacity (bold line) and temperature(thin
line) versus $r_{+}$ for $q=1$, $P=1$ $r_{0}=1$ and $z=0$ (continuous green
line), $z=0.5$ (dotted black line) and $z=0.8$ (dashed blue line). }
\label{Fig2}
\end{figure}

According to Fig. 2, we find that the heat capacity has no divergence point
which is due to the fact that $\mathcal{B}$\ is a positive definite
function. In addition, one can find a real positive root for $\mathcal{A}=0$
called bound point $r_{0}$. We should note that $r_{0}$ is the root of
temperature, simultaneously, and both $T$ and $C_{P,Q}$\ are positive
definite functions for $r_{+}>r_{0}$. Roughly speaking, the obtained
solutions are thermally stable.

\subsection{ Smarr relation}

The Smarr relation \cite{Smarr} as one of thermodynamical relation
in the case of black holes has attracted attention. In this
section, we derived this relation for black holes in three
dimensional with Lifshitz-like spacetime. Smarr relation, together
with the first law of black hole thermodynamics \cite{Bardeen} has
a main role in black hole physics. The variable cosmological
constant is considered as a thermodynamic pressure to explain
scaling relation of Smarr formula \cite{Creighton,
Caldarelli,Dolan2,Cvetic,Gunasekaran}. Taking into account the
scaling argument for our Lifshitz like solutions in the extended
phase
space, one can find the following Smarr relation is hold%
\begin{equation}
\gamma M=\left( 1+\gamma -\frac{z}{2}\right) TS+\frac{\delta }{3}\Phi Q-2PV,
\label{SMARR}
\end{equation}%
where $T=\left( \frac{\partial M}{\partial S}\right) _{Q}$ is temperature as
calculated in Eq. (\ref{Temp2}) and $\Phi $ is a modified potential that
calculated at the event horizon of Lifshitz like black hole solutions as%
\begin{equation}
\Phi =\left( \frac{\partial M}{\partial Q}\right) _{S}=\frac{q}{r_{+}}\left(
\frac{r_{+}}{r_{0}}\right) ^{1+\gamma -\delta }.  \label{PO}
\end{equation}%
It is notable that for $z=0$, Eq. (\ref{SMARR}) reduces to that of
nonlinearly charged BTZ black holes in which mass term has no scaling.

\section{Conclusion}

In this paper, we have obtained a new Lifshitz-like charged black hole
solutions in three dimensional pure $F(R)$ gravity. We have discussed
geometrical properties of the solutions and found that these solutions
reduce to charged BTZ like solutions in Einstein-$\Lambda $-PMI gravity.

We have also investigated thermodynamic quantities of black holes and showed
that these quantities satisfy the first law of thermodynamics. Next, we look
for possible phase transition in the extended phase space thermodynamics and
found that the three dimensional Lifshitz-like black holes do not have any
critical behavior. Then, we studied thermal stability and calculated the
heat capacity of the Lifshitz-like black hole solutions in canonical
ensembles and found that regarding the root of the temperature as a bound
point ($r_{0}$), the obtained solutions are thermally stable for $%
r_{+}>r_{0} $. Finally, we obtain modified Smarr relation in the presence of
Lifshitz parameter and found that regardless of the cosmological constant
term, the scaling of other thermodynamic quantities is modified.

\section{Appendix}

Here, we obtain the higher dimensional topological black hole
solutions in a Lifshitz-like background spacetime. Thus, we
consider the metric of $d$-dimensional spacetime as
\begin{equation}
ds^{2}=-(\frac{r}{r_{0}})^{z}B(r)dt^{2}+\frac{dr^{2}}{B(r)}+r^{2}d\Omega
^{2},  \label{Metd}
\end{equation}%
where
\begin{equation*}
d\Omega ^{2}=\left\{
\begin{array}{cc}
dx_{1}^{2}+\sum\limits_{i=2}^{d-2}\prod\limits_{j=1}^{i-1}\sin
^{2}x_{j}dx_{i}^{2}, & k=1 \\
dx_{1}^{2}+\sinh ^{2}x_{1}\left(
dx_{2}^{2}+\sum\limits_{i=3}^{d-2}\prod\limits_{j=2}^{i-1}\sin
^{2}x_{j}dx_{i}^{2}\right) , & k=0 \\
\sum\limits_{i=1}^{d-2}dx_{i}^{2}, & k=-1 \\ & \\ &
\end{array}%
\right. .
\end{equation*}

Considering Eq. (\ref{Metd}), we can extract the metric function for $R=R_{0}
$, where
\begin{equation}
R_{0}=-B^{\prime \prime }-\frac{3z+4d-8}{2r}B^{\prime }
-\frac{z^{2}+2z(d-3)+(d-2)(d-3)}{2r^{2}}B
+\frac{k(d-2)(d-3)}{r^{2}},
\end{equation}%
with the following exact solutions
\begin{equation}
B(r)=K-\frac{m}{r^{\gamma }}+\frac{q^{2}}{r^{\delta }}-\Lambda r^{2},
\label{sol-d}
\end{equation}%
in which $K$ is related to the horizon topology and $\Lambda $ is
a (positive/negative or zero) constant that its value is depending
on the
sign/value of $R_{0}$ with the following explicit forms%
\begin{eqnarray*}
K &=&\frac{2(d-2)(d-3)k}{z^{2}+2z(d-3)+2(d-2)(d-3)}, \\
\Lambda  &=&\frac{2R_{0}}{z^{2}+2dz+2d(d-1)}.
\end{eqnarray*}%
In addition, $m$ and $q$ are two integration constants while $\gamma $ and $%
\delta $ are defined as%
\begin{eqnarray*}
\gamma  &=&\frac{1}{4}\left( 3z+4d-10-\sqrt{z^{2}+(8d-12)z+4}\right) , \\
\delta  &=&\frac{1}{4}\left( 3z+4d-10+\sqrt{z^{2}+(8d-12)z+4}\right) .
\end{eqnarray*}

\begin{acknowledgements}
The authors wish to thank the anonymous referee for the
constructive comment. SHH would like to thank Shiraz University
Research Council.
\end{acknowledgements}

\end{document}